\documentclass[aps,twocolumn,showpacs,superscriptaddress,nofootinbib]{revtex4}

\usepackage{graphicx}
\usepackage{latexsym}

\usepackage{color}

\begin{document}
\title{
Complementary mode analyses between sub- and super-diffusions
}

\author{Takuya Saito}
\email[Electric mail:]{tsaito@eri.u-tokyo.ac.jp}
\affiliation{Earthquake Research Institute, University of Tokyo, Tokyo 113-0032, Japan}

\author{Takahiro Sakaue}
\email[Electric mail:]{sakaue@phys.kyushu-u.ac.jp}
\affiliation{Department of Physics, Kyushu University, Fukuoka 819-0395, Japan}
\affiliation{JST, PRESTO, 4-1-8 Honcho Kawaguchi, Saitama 332-0012, Japan}

\def\Vec#1{\mbox{\boldmath $#1$}}
\def\degC{\kern-.2em\r{}\kern-.3em C}

\def\SimIneA{\hspace{0.3em}\raisebox{0.4ex}{$<$}\hspace{-0.75em}\raisebox{-.7ex}{$\sim$}\hspace{0.3em}} 

\def\SimIneB{\hspace{0.3em}\raisebox{0.4ex}{$>$}\hspace{-0.75em}\raisebox{-.7ex}{$\sim$}\hspace{0.3em}}

\date{\today}

\begin{abstract}
Several sub-diffusive stochastic processes in nature, e.g., motion of tagged monomer in polymers, height fluctuation of interfaces and particle dynamics in single-file diffusion etc. can be described rigorously or approximately by the superposition of various modes whose relaxation times are broadly distributed. 
In this paper, we propose a mode analysis generating super-diffusion, which is paired or complementary with that for sub-diffusion.
The key point in our discussion lies in the identification of a pair of conjugated variables, which undergoes sub- and super-diffusion, respectively. We provide a simple interpretation for the sub- and superdiffusion duality for these variables using the language of polymer physics. The analysis also suggests the usefulness to look at the force fluctuation in experiments, where a polymer is driven by a constant velocity.
\end{abstract}


\def\degC{\kern-.2em\r{}\kern-.3em C}

\newcommand{\gsim}{\hspace{0.3em}\raisebox{0.5ex}{$>$}\hspace{-0.75em}\raisebox{-.7ex}{$\sim$}\hspace{0.3em}} 
\newcommand{\lsim}{\hspace{0.3em}\raisebox{0.5ex}{$<$}\hspace{-0.75em}\raisebox{-.7ex}{$\sim$}\hspace{0.3em}} 

\def\Vec#1{\mbox{\boldmath $#1$}}

\maketitle

\section{Introduction}
There has been an increasing number of reports on stochastic processes~\cite{Mandelbrot,Amitai,Krug,PhysRep_Metzler_2000,AdvPhys_Havlin_2002,PhysToday_Barkai_2012,PRL_Amblard_1996,PRL_Wong_2004,PRL_Akimoto_2011,PRL_Jeon_2012,NewJPhys_Godec_2014,JPhysChemLett_Xue_2016,PRL_Bronstein_2009,BiophysJ_Levi_2005,NuclAcidsRes_Kuwada_2013,BJ_Lampo_2015,PRL_Weber_2010,PLoS_Shinkai_2016,PRE_Vandebroek_2015,SoftMatter_Sakaue_2016,PRE_Lizana_Barkai_Lomholt_2010,BRL_Ooshida_2016,JCP_Panja_2009,JStatMech_Keesman_Panja_2013,EPL_Mizuochi_2014}, where the mean square displacement (MSD) of the observable $x(t)$ grows nonlinearly with time.
\begin{eqnarray}
\langle \Delta x_t^2 \rangle \sim t^{\alpha}  \qquad (\alpha \neq 1),
\label{MDS_alpha}
\end{eqnarray}
where $\Delta x_t \equiv x(t_0+t) - x(t_0)$ is the displacement during the time interval $t$, and the bracket indicates appropriate averaging.

This so-called anomalous diffusion has been found in various systems that have generated interest not only in physics~\cite{Amitai,Krug,PhysRep_Metzler_2000,AdvPhys_Havlin_2002,PhysToday_Barkai_2012,PRL_Amblard_1996,PRL_Wong_2004,PRL_Akimoto_2011,PRL_Jeon_2012,NewJPhys_Godec_2014,JPhysChemLett_Xue_2016,PRL_Bronstein_2009,BiophysJ_Levi_2005,NuclAcidsRes_Kuwada_2013,BJ_Lampo_2015,PRL_Weber_2010,PLoS_Shinkai_2016,PRE_Vandebroek_2015,SoftMatter_Sakaue_2016,PRE_Lizana_Barkai_Lomholt_2010,BRL_Ooshida_2016,JCP_Panja_2009,JStatMech_Keesman_Panja_2013,EPL_Mizuochi_2014}, but also in other disciplines~\cite{PRL_Bronstein_2009,BiophysJ_Levi_2005,NuclAcidsRes_Kuwada_2013,BJ_Lampo_2015,PRL_Weber_2010,PLoS_Shinkai_2016}. Many intriguing examples can be observed in the dynamics of complex fluids~\cite{PRL_Amblard_1996,PRL_Wong_2004,PRL_Akimoto_2011,PRL_Jeon_2012,NewJPhys_Godec_2014,JPhysChemLett_Xue_2016}. Intracellular transport is a related realm in out-of-equilibrium systems, which is under active investigation in current biophysics studies~\cite{PhysToday_Barkai_2012,PRL_Bronstein_2009,BiophysJ_Levi_2005,NuclAcidsRes_Kuwada_2013,BJ_Lampo_2015,PRL_Weber_2010,PLoS_Shinkai_2016,PRE_Vandebroek_2015,SoftMatter_Sakaue_2016}.
A time series of stock prices is yet another example, which enters the long list of anomalous diffusion phenomena.

Often (but not always), the anomaly $\alpha \neq 1$ indicates the presence of spatial or temporal long range correlation~\cite{Mandelbrot,Krug}, which produces a memory effect for the motion of the observable. From the standpoint of physics, one is then interested in the mechanism, through which such a memory is constructed from the microscopic model.
One of the few problems, in which the program of the coarse-graining can be done rigorously, is a tagged monomer diffusion in the Rouse polymer, for which $\alpha = 1/2$~\cite{deGennesBook,Doi_Edwards,JStatMech_Panja_2010,PRE_Sakaue_2013}\footnote{As related problems, one may also cite the height fluctuation of interface~\cite{Krug} and single-file diffusion~\cite{PRE_Lizana_Barkai_Lomholt_2010,BRL_Ooshida_2016}.}. Here, as we shall review shortly, one can derive the generalized Langevin equation (GLE) for the motion of the tagged monomer~\cite{JStatMech_Panja_2010,PRE_Sakaue_2013,PRE_Saito_2015}. Through the derivation, one can clearly see that the long-range temporal memory is built up by the superposition of modes, whose relaxation times are broadly distributed. One can also find an idea on how to ``tune" the diffusion exponent $\alpha$ by customizing the mode spectrum and the weight of the mode superposition~\cite{Amitai,Krug}. This paves the way for approximate schemes to include various nonlinear effects in polymer dynamics such as the hydrodynamic interaction (the Zimm model) and the excluded-volume interaction~\cite{deGennesBook,Doi_Edwards,JStatMech_Panja_2010,PRE_Sakaue_2013,PRE_Saito_2015,Grosberg_Khokhlov}. From more general viewpoint, this provides an approach to design a microscopic model for the anomalous diffusion with the desired exponent $\alpha$.

It turns out, however, that the simple generalization of the Rouse model for the tunable exponent $\alpha$ is able to produce the sub-diffusion $0<\alpha < 1$ only~\cite{Amitai,Krug}. One is then led to the following question; what is the corresponding scheme for the super-diffusion $1<\alpha <2$ ?

We address this question by taking the Rouse model as a paradigmatic example. The key point in our discussion lies in the identification of the pair of conjugate stochastic variables, say the position $x(t)$ and the momentum $p(t)$ in such a way that when one of the variables (say, $x$) performs the sub-diffusion, then the other exhibits the super-diffusion, the MSD of which for the Rouse model is indeed $\langle \Delta p_t^2 \rangle \sim t^{\alpha_p}$ with $\alpha_p = 3/2$. 
This allows us to establish the paired correspondence, or mutually complementary relations between respective equations of motion in mode space. For sub-diffusion case, the general structure of the mode equation is the force balance between the frictional force and the restoring force with wavelength-dependent spring constant~\cite{Doi_Edwards,Amitai,JStatMech_Panja_2010,PRE_Sakaue_2013,PRE_Saito_2015}.
On the other hand, we will see for super-diffusion that, 
virtually, ``fictive inertia" is balanced by the ``fictive friction"
with the wavelength-dependent coefficient.
We also point out that this super-diffusive analysis fits naturally in the single polymer manipulation experiment, where the polymer is driven by a constant velocity.

\section{Pair of conjugated variables and protocols}
\label{pair_variables}
We consider the situation, in which the force $f(t)$ is acting on a tagged monomer in the Rouse chain. Let $x(t)$ denote the position of the tagged monomer. Note that, throughout this paper. we suppress the vector notation, which is not essential for our discussion. 
Let us define a variable $p(t)$ such that $dp(t)/dt = f(t)$. Thus $p(t)$ is the impulse or the momentum transferred to the tagged monomer, and identified as a conjugate stochastic variable to $x(t)$. 

Along with these conjugated stochastic variables, it is useful to identify a pair of conjugated protocols with which the polymer is manipulated.
In the {\it force control} protocol, the external force $f_{ext}(t)$ is imposed on the tagged monomer. The most basic protocol is to apply the step force $f_{ext}(t)= f \Theta (t)$ with $f$ or $\Theta (t)$ denoting the force magnitude or the Heaviside step function, respectively. 
On the other hand, in the {\it velocity control} protocol, the tagged monomer is moved according to externally imposed velocity $v_{ext} (t)$. In the simplest case, one starts to move the tagged monomer at constant velocity, such that the imposed velocity of the tagged monomer is $v_{ext}(t) = v\Theta (t)$. Here, the force $f(t)$, thus $p(t)$, too, are fluctuating variables.

The GLE for the anomalous dynamics of the tagged monomer can be written as a time evolution of either one of conjugated variables $x(t)$ or $p(t)$; 
\begin{eqnarray}
\frac{dx(t)}{dt}&=&\int_{-\infty}^t ds \ \mu (t-s)f_{ext}(s)+\delta v (t),
\label{GLE_f}
\\
&\Updownarrow& \nonumber 
\\
\frac{dp(t)}{dt}&=&\int_{-\infty}^t ds \ \Gamma (t-s)v_{ext}(s)+\delta f (t),
\label{GLE_v}
\end{eqnarray}
where we employ the force or velocity control protocol for the equation of $x(t)$ or $p(t)$, respectively.  
Thus, in Eq.~(\ref{GLE_f}), $f_{ext}(t)$ is an externally imposed quantity, and $\int_0^t ds \ \mu (s)$ describes the response to the force $f_{ext}(t)= f \Theta (t)$, i.e., the average velocity $\langle v(t) \rangle = d\langle x(t)\rangle/dt$ divided by $f$, of the pulled monomer.
Similarly, Eq.~(\ref{GLE_v}) states that $\int_0^t ds \Gamma (s) = (d\langle p(t) \rangle/dt) /v =\langle f(t) \rangle /v$ describes how the average force builds up with time in response to the imposed velocity $v_{ext}(t) = v\Theta (t)$. 
In thermal system, the memory kernel $\mu (t)$, $\Gamma (t)$ in respective equation is related to the noise $\delta v (t)$, $\delta f (t)$ via fluctuation-dissipation theorem (FDT); $k_BT\mu (t-s)=\left< \delta v(t)\delta v(s) \right>\label{2FDR_f}$, $k_BT \Gamma (t-s)=\left< \delta f (t)\delta f (s) \right>\label{2FDR_v}$ with $k_B T$ being the thermal energy.

As we will review in Sec.~\ref{Mode_sub}, Eq.~(\ref{GLE_f}) can be derived from a set of microscopic equations of motion by eliminating all the degrees of freedom except for the tagged monomer~\cite{JStatMech_Panja_2010,PRE_Sakaue_2013,PRE_Saito_2015}. For the Rouse chain, the memory kernel is calculated as $\mu(t) \sim - t^{-3/2}$. It is possible to tune the microscopic equations of motion to customize the kernel as $\mu(t) \sim - t^{\alpha_x-2}$ with $0<\alpha_x<1$. In the absence of the external force $f_{ext}(t)=0$, the position of the tagged monomer is kicked by the fractional noise $\delta f(t)$. Then, the sub-diffusion scaling of MSD 
\begin{eqnarray}
\left< \Delta x_t^2 \right> \sim t^{\alpha_x}   \label{Dx2}
\end{eqnarray}
for $x(t)$ is obtained by using FDT.
In more general case with the driving force $f_{ext}(t)$, the same scaling holds for the variance of the displacement
\begin{eqnarray}
\left< [\delta (\Delta x_t)]^2 \right> = \left< \Delta x_t^2 \right>- \left< \Delta x_t \right>^2 \sim t^{\alpha_x},   \label{Dx2}
\end{eqnarray}
where $\delta (\Delta x_t) = \Delta x_t - \left< \Delta x_t \right>$.
Now, comparing Eqs~(\ref{GLE_f}) and~(\ref{GLE_v}), one recognizes that the mobility kernel $\mu(t)$ is the ``inverse" of the friction kernel $\Gamma(t)$. More precisely, ${\hat \mu}(y) {\hat {\Gamma}}(y)=1$, where we introduce the Laplace transform of a function $f(t)$ as ${\hat f}(y) = \int_0^{\infty} dt \ f(t) e^{-yt}$. This indicates the correspondence $\mu(t) \sim -  t^{\alpha_x-2}  \Leftrightarrow \Gamma(t) \sim t^{-\alpha_x}$ in the time domain. Again, using FDT, we find the super-diffusion scaling for MSD (more generally, the variance) of the momentum as
\begin{eqnarray}
\left<[\delta (\Delta p_t)]^2 \right> \sim t^{\alpha_p}    \label{Dp2}
\end{eqnarray}
with $\alpha_p=2 - \alpha_x$.
In Sec.~\ref{Mode_super}, we propose the mode analysis for this super-diffusive dynamics.

\section{Mode analysis of sub-diffusion}
\label{Mode_sub}
\subsection{Rouse model}
\label{Mode_sub_Rouse}
Let us consider a linear polymer with $N+1$ monomers under the force control protocol. The monomers are labeled by the index $n$ from one end and the position of $n$-th monomer is $x_n(t)$.
In the Rouse model, the neighboring monomers are connected by a harmonic spring. The equation of motion in the continuum limit takes the following form of noisy diffusion equation~\cite{deGennesBook,Doi_Edwards,Grosberg_Khokhlov} 
\begin{eqnarray}
\gamma \frac{\partial x_n}{\partial t} &=& k \frac{\partial^2 x_n}{\partial n^2} +f_n(t),
\label{Rouse}
\end{eqnarray}
where $k$, $\gamma$ are the spring constant and the frictional coefficient per monomer, respectively. The first and the second cumulants of the force characterize the external force $\langle f_n(t) \rangle$ and the noise strength $\langle \delta f_n(t) \delta f_m(s) \rangle = 2 \gamma k_BT \delta(t-s) \delta(n-m)$. The external force is acting on the tagged monomer only with the label $n=n_0$, $\langle f_n(t)\rangle=f_{ext}(t)\delta_{nn_0}$.

By imposing the open boundary condition at both chain ends, we introduce the normal coordinate
\begin{eqnarray}
X_q(t)=\int_0^N dn \ h_{q,n} \ x_n(t) \qquad (q=0, 1, 2, \cdots),
\label{normal_coordinate}
\end{eqnarray}
with
\begin{eqnarray}
h_{q,n}= \frac{1}{N}\cos{\left( \frac{\pi q n}{N}\right)}
\end{eqnarray}
Note that while $X_0(t)$ is the center of mass, other modes $X_q(t)$ with $q \ge 1$ represent the internal deformation dynamics. 
In the normal mode space, the Hamiltonian of the Rouse model is diagonalized, and each mode evolves independently according to the equation for the mode number $q$~\cite{Doi_Edwards,JCP_Panja_2009};
\begin{eqnarray}
\gamma_q\frac{dX_q}{dt} &=& -k_q X_q +F_q(t),
\label{Xq_eq}
\end{eqnarray}
which takes a form of the overdamped Langevin equation of a particle trapped in a harmonic potential. Here the force $f_n(t)$ is transformed to $F_q(t)$ by the same formula as $x_n(t)$ (Eq.~(\ref{normal_coordinate})), and 
\begin{eqnarray}
\gamma_q=\gamma, \quad k_q=k\left(\frac{\pi q}{N} \right)^2,
\label{k_q_Rouse}
\end{eqnarray}
The noise correlation in the normal coordinate space reads~\footnote{There are some arbitrariness for the definition of $\gamma_q$ and $k_q$ (with the relaxation time $\gamma_q/k_q$ being fixed) and the transformation rule of the force, which affects the proportionality factor in the normal mode space FDT. While, in Refs.~\cite{PRE_Sakaue_2013,PRE_Saito_2015}, we adopted the convention in the textbook~\cite{Doi_Edwards}, here we adopt a different convention, which would be more natural evoking standard Fourier analysis. This avoid a factor $\sim N$ in the transformation formula of the force. }
\begin{eqnarray}
\langle \delta F_q(t) \delta F_{q'}(t') \rangle =  (\gamma_q/N) (1+\delta_{q0}) k_BT \delta_{q q'} \delta(t-t').
\label{FDT_F_q}
\end{eqnarray}
The first moment of $F_q(t)$ is proportional to the external force acting on the tagged monomer, i.e., $\langle F_q(t) \rangle = h_{q,n_0}f_{ext}(t)$, which is not a fluctuating quantity.

Equation~(\ref{Xq_eq}) is exactly solved and the motion of the tagged monomer $x(t) \equiv x_{n_0}(t)$ is given through the inverse transformation of Eq.~(\ref{normal_coordinate}) as
\begin{eqnarray}
x (t)
&=& 
\sum_{q \ge 0} X_q(t) h^{\dagger}_{q,n_0}
\label{x_solution}
 \\
&=& 
X_0(t)+\sum_{q \ge 1}
\int_{-\infty}^t ds \ \frac{F_q(s)}{\gamma_q} e^{-(k_q/\gamma_q)(t-s)} h^{\dagger}_{q,n_0},
\nonumber 
\end{eqnarray}
where
\begin{eqnarray}
h_{q,n}^\dagger = 2\frac{\cos{\left(\frac{ \pi  n q}{N} \right)}}{1+\delta_{q0}}.
\end{eqnarray}
Upon time derivative, Eq.~(\ref{x_solution}) can be arranged in the form of Eq.~(\ref{GLE_f}) with FDT, where an explicit expression for the memory kernel is
\begin{eqnarray}
\mu(t) = &&\mu_{cm}(t)+\sum_{q\ge1} \frac{1}{N\gamma_q}\delta(t) (h^{\dagger}_{q,n_0})^2 \nonumber \\
 &-& \sum_{q\ge1} \frac{k_q}{2N \gamma_q^2}e^{-(k_q/\gamma_q)t} (h^{\dagger}_{q,n_0})^2
\label{mu_Rouse}
\end{eqnarray}
Here $\mu_{cm}= 2/(N \gamma) \delta(t)$ represents the contribution of the center-of-mass mode.
In addition to the instantaneous response, i.e., the second term $\simeq 2 \gamma^{-1} \delta (t)$, one can clearly see the presence of persistent memory, which is built from the superposition of many modes with broadly distributed relaxation times.
In the time window $ \tau_u \ll t \ll \tau_u N^2$, the summation can be approximated by the Gaussian integral, where $\tau_u \equiv \gamma/k$ is the shortest time scale in the model. This yields a power-law memory $\simeq -(\tau_u \gamma)^{-1} |t/\tau_u|^{-3/2}$ with negative sign, as already announced in Sec.~\ref{pair_variables}, hence, $\alpha_x=1/2$ for the Rouse polymer.

\subsection{Sub-diffusion with tunable exponent}
\label{tunable_exponent_subdiffusion}
It is possible to control the subdiffusion exponent $\alpha_x$ by tuning the spectrum of mode distribution. This can be done by modifying the wavenumber-dependence of spring constant $k_q$ as in Ref.~\cite{Krug} or the friction constant $\gamma_q$ as in Ref.~\cite{Amitai}. Here, we present somewhat generalized argument using a language of polymer physics.
Let us define the following scaling forms
\begin{eqnarray}
k_q &\simeq& k \left(\frac{q}{N}\right)^{1+2\nu} \label{k_q_general}\\
\gamma_q &\simeq& \gamma  \left( \frac{q}{N}\right)^{1-(z-2)\nu} \qquad (q \ge 1)
\label{gamma_q_general}
\end{eqnarray}
for the spring and the friction constants in the mode equation in Eq.~(\ref{Xq_eq})~\cite{Doi_Edwards,JCP_Panja_2009}. For $q=0$ mode, there is no restoring force $k_q=0$ and one may assume $\gamma_0 \simeq \gamma_1$ such that the translational diffusion time $\sim N^{2\nu}/D_{cm}$ with $D_{cm} \simeq k_BT/(\gamma_0 N)$ has the same scaling form as that of the longest relaxation time $\gamma_1/k_1$. 
This leads to the equilibrium size $R_m$ of the sub-chain with $m=N/q$ monomers 
\begin{eqnarray}
R_m^2 \equiv \langle (r_{m_0+m}(t) - r_{m_0}(t))^2 \rangle \simeq \frac{k_BT}{k} m^{2\nu}
\end{eqnarray}
and the relaxation time $\tau_m$ of the corresponding section\footnote{Thus, $\nu$ and $z$ are, respectively, static and dynamic exponents familiar in the context of critical phenomena. For clarity, we note the relation with Refs.~\cite{Krug} and~\cite{Amitai}: In Ref.~\cite{Krug}, the {\it free draining} dynamics $z=2 + \nu^{-1}$ is assumed, and the restoring force is controlled by the exponent $z'=1+2\nu$, hence $k_q \sim q^{z'}, \ \gamma_q \sim q^{0}$. On the other hand, in Ref.~\cite{Amitai}, the nearest-neighbor harmonic spring interaction, thus, $\nu=1/2$ is assumed, while the mobility is subjected to the exponent $z'' = z/2$, hence $k_q \sim q^2, \ \gamma_q \sim q^{2-z''}$}:
\begin{eqnarray}
\tau_m \simeq \tau_u (N/q)^{\nu z} \simeq \tau_u (R_m)^z
\end{eqnarray}
The analysis in Sec.\ref{Mode_sub_Rouse} up to Eq.~(\ref{mu_Rouse}) is intact, but now the time dependence in the memory term is different due to the change in the mode spectrum. Using a formula 
\begin{eqnarray}
\int_0^{\infty}dx \ x^{b-1}e^{-a x^{\theta}} = \Gamma(b/\theta)/(\theta a^{b/\theta})
\label{integral_formula}
\end{eqnarray}
 for $a,b, \theta >0$ (the symbol $\Gamma (\cdot)$ here is used for the Gamma function and should not be confused with the friction kernel), which may be viewed as a generalization of the Gaussian integral, one finds the following power-law memory
\begin{eqnarray}
\mu(t) - \mu_{cm}(t) - 2 \gamma^{-1}\delta(t)  \simeq  - \frac{1}{\tau_u \gamma} \left|\frac{t}{\tau_u}\right|^{-2 + (2/z)}.
\label{mu_sub_general}
\end{eqnarray}
The required condition $b>1$ for this integral formula seems to imply that Eq.~(\ref{mu_sub_general}) would be valid for $z>1$. However, the sum-rule~\cite{PRE_Saito_2015}
\begin{eqnarray}
{\hat \mu}(0)= \int_0^{\infty} dt \ \mu(t) = \frac{1}{\gamma N},
 \label{sum-rule}
\end{eqnarray}
as verified from Eq.~(\ref{mu_Rouse}), indicates stronger condition $z>2$. Physically, this sum-rule reflects the fact that, in the long time limit, i.e., time scale longer than the longest relaxation time $\simeq \tau_u N^{\nu z}$ in the problem, only the center-of-mass mode survives.
For $z>2$, the time integral of $\mu(t)$ from Eq.~(\ref{mu_sub_general}) can be evaluated as
\begin{eqnarray}
\int_0^{\infty}dt \ \mu(t) && 
\simeq
 \frac{1}{\gamma N} + \frac{1}{\gamma}  \nonumber \\
&-& \frac{z}{(z-2)\gamma}( \tau_u)^{1-(2/z)} \epsilon^{-1+(2/z)}  
\label{mu_integral}
\end{eqnarray}
where the lower cut-off $\epsilon \sim \tau_u$ is associated with the shortest time scale in the model. The sum-rule Eq.~(\ref{sum-rule}) is ensured in this way. However, the form of Eq.~(\ref{mu_sub_general}) with $z \le 2$ does not satisfy the sum-rule, and therefore is invalid.

The MSD can be calculated via FDT as
\begin{eqnarray}
\langle [\delta(\Delta x_t)]^2 \rangle = k_BT \int_0^t dt_1 \int_0^t  dt_2 \ \mu(t_1-t_2).
\label{MSD_from_mu}
\end{eqnarray}

For $z>2$, the power-law part of the above double integral can be evaluated as
\begin{eqnarray}
&&- \frac{1}{\tau_u \gamma}\int_0^t dt_1 \int_0^t  dt_2 \ \left|\frac{t_1-t_2}{\tau_u} \right|^{-2 + (2/z)} \nonumber \\
 &\simeq& - 
\frac{2}{\gamma} 
 \int_0^t dt_1 \int_{\epsilon/\tau_u}^{t_1/\tau_u}  ds \ s^{-2+(2/z)} \nonumber \\
&=& c_1 (t/\tau_u)^{2/z} - c_2 t/\tau_u,
\label{double_integral}
\end{eqnarray}
where $c_1 = z^2 \tau_u/[\gamma (z-2)] >0$ and $c_2 = [2z\tau_u/\gamma (z-2)](\epsilon/\tau_u)^{-1+(2/z)} >0$. The term $\sim -c_2 t$ cancels the contribution to MSD from the instantaneous response $2 \gamma^{-1} \delta(t)$ in Eq.~(\ref{mu_sub_general}).
In this way, the sub-diffusive MSD exponent $\alpha_x = 2/z$ ($z>2$) is obtained, but there is no way to generate super-diffusion along the present line of argument.

In Fig.~1, we plot MSD calculated according to Eq.~(\ref{MSD_from_mu}) for various exponent $z$, where the mobility kernel is given by discrete sum as Eq.~(\ref{mu_Rouse}) with the contribution of center-of-mass mode subtracted.
As the above discussion indicates, the sub-diffusion exponent $\alpha_x = 2/z$ is successfully reproduced for $z>2$, but smaller values of $z$ do not. The latter cases seem only to become close to the normal diffusion $\alpha_x=1$ behavior in the long time limit, 
suggesting that MSD of $x(t)$ generated by eq.~(\ref{Xq_eq}) displays the sub-diffusion $0<\alpha_x<1$ only.

\begin{figure}[tb]
\begin{center}
\includegraphics[scale=0.50]{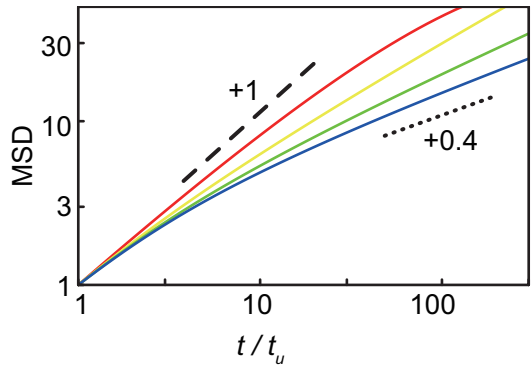}
      \caption{
	  (Color Online) 
	  Plot of MSD calculated according to Eq.~(\ref{MSD_from_mu}) for $z=3/2$ (red), $3$ (yellow), $4$ (green), $5$ (blue), where the mobility kernel is given by discrete sum as Eq.~(\ref{mu_Rouse}). The chain length is $N=10^3$ and the contribution of center-of-mass mode is subtracted.
	  Dashed line represents the linear growth, which is the upper bound of the slopes generated by eq.~(\ref{Xq_eq}).
	  Dotted line shows the the theoretically predicted asymptotic slope in the case of $z=5$ in the long time limit.
	  All the plots are normalized as $\mathrm{MSD}(t_u)=1$. 
	  }
\label{subdiff}
\end{center}
\end{figure}

\section{Mode analysis of super-diffusion}
\label{Mode_super}
How can we construct a super-diffusion as a superposition of various modes? To answer this, let us recall the relation between conjugated variables defined in Sec.~\ref{pair_variables}.
We then realize that our task is to find a way of how to decompose eq.~(\ref{GLE_v}) into the mode $P_q(t)=\int_q dn h_{q,n} p_n(t)$.
To clarify the argument, we again take the Rouse model in this section.

In Sec.~\ref{Mode_sub_Rouse}, we have analyzed the Rouse model in the constant force protocol and constructed the power-law mobility kernel $\mu(t)$, valid in the time scale $\gamma/k \ll t \ll (\gamma/k)N^2$, by the superposition of normal modes (Eq.~(\ref{mu_Rouse})).
Here, let us consider a constant velocity protocol, in which the labeled monomer gets driven by a constant velocity $v_{ext}(t) = v \Theta (t)$.
Now the position $x(t) = vt $ is not fluctuating, instead the force $f(t)$ acting on the labeled monomer is a fluctuating quantity. 
The power-law form of the mobility kernel and the relation ${\hat \Gamma}(y) {\hat \mu}(y)=1$ leads to the following power-law form of the friction kernel
\begin{eqnarray}
\Gamma(t) \simeq +k \left|\frac{t}{\tau_u}\right|^{-1/2} 
\label{Gamma_power_law_Rouse}
\end{eqnarray}
which is again valid in the time scale $\tau_u \ll t \ll \tau_uN^2$~\footnote{In long time scale $t > \tau_u N^2$, the memory decays exponentially, which implies ${\hat \Gamma} (0) \simeq \int_0^{\tau_u N^2} dt \  \Gamma (t) \simeq \gamma N$ consistent with the sum-rule (Eq.~(\ref{sum-rule})).
In the short time limit $\Gamma (t \rightarrow \tau_u) \simeq \gamma/\tau_u$, knowing that $\tau_u$ is the shortest time scale in the problem, we may identify $\Gamma(t) = 2\gamma \delta (t)$ in agreement with physical intuition.}.

Now, taking the reverse direction from Eq.~(\ref{mu_Rouse}) to power-law mobility kernel, we decompose the power-law frictional kernel in the following way:
\begin{eqnarray}
\Gamma(t)
&\simeq &
 \sum_{q\ge 1}  \ 
 \frac{Nk_q}{q^2} e^{-\frac{k_q}{\gamma_q}t}
(h_{q,n_0}^\dagger)^2 
\label{decomp_Gamma}
\end{eqnarray}
Then, the GLE (Eq.~(\ref{GLE_v})) indicates the decomposition of the average force into the sum of independent modes $F_q(t)$:
\begin{eqnarray}
\left< f(t) \right> = \sum_{q \ge 0} \langle F_q(t) \rangle h^{\dagger}_{q,n_0} 
\end{eqnarray}
with
\begin{eqnarray}
\langle F_q(t) \rangle &\simeq&  vN  \int_0^t ds \ 
 \frac{k_q}{q^2} e^{-\frac{k_q}{\gamma_q}(t-s)} h^{\dagger}_{q,n_0} 
\nonumber \\
 &=& v \frac{N\gamma_q}{q^2}[1-e^{-\frac{k_q}{\gamma_q}t}] h^{\dagger}_{q,n_0}
 \qquad (q \geq 1).
\label{decomp}
\end{eqnarray}
This suggests that the average time evolution of individual modes for $q \geq 1$ are governed by
\begin{eqnarray}
m_q
\frac{d \left< F_q(t) \right>}{dt}= -g_q\left< F_q(t) \right> +\langle V_q(t) \rangle
\label{average_Fq}
\end{eqnarray}
where 
\begin{eqnarray}
m_q =  k_q^{-1} (q/N)^2 \simeq k^{-1} , \ g_q = \gamma_q^{-1} (q/N)^2 = \gamma^{-1} (q/N)^{2}
\label{k_q_tilde}
\end{eqnarray}
 and $\langle V_q(t) \rangle = \int_0^N  h_{q,n} v_{ext}(t) \delta_{n n_0} = h_{q,n_0}v \Theta(t) \simeq v \Theta(t)h^{\dagger}_{q,n_0}/N$.
Equation of motion for $q=0$ is obtained from the force balance $F_0(t)= (g_0)^{-1}V_0(t)$, which is reproducible by putting $m_{0}=0$ and $g_{0}=g_{1}$ into eq.~(\ref{average_Fq}).

To connect this average dynamics to diffusion, we generalize Eq.~(\ref{average_Fq}) to the stochastic differential equation
\begin{eqnarray}
m_q \frac{d^2 P_q(t)}{dt^2} 
&=& 
-g_q \frac{dP_q(t)}{dt}
+ V_q(t),
\label{Lang_P}
\end{eqnarray}
where, to make the correspondence with the mode equation~(\ref{Xq_eq}) for $X_q(t)$ clear, 
we rewrite Eq.~(\ref{average_Fq}) in terms of $P_q(t) \equiv P_q(0) +\int_0^t ds \ F_q (s)$, and the noise $\delta V_q(t) = V_q(t) - \langle V_q(t) \rangle$ is to be appropriately related to the dissipative response via FDT.
Equation~(\ref{Lang_P}) resembles the underdamped Langevin equation for a small particle with ``mass " $m_q$ and ``friction constant" $g_q$ in the Newtonian fluid.
Indeed, one can verify that the noise correlation of the form 
\begin{eqnarray}
 \left< \delta V_q(t) \delta V_{q'}(t') \right> = \frac{2k_BTg_q}{N} \delta_{qq'} \delta(t-t'),
 \label{FDT_V_q}
\end{eqnarray}
which is suggested by such a particle analogy (see Eq.~(\ref{FDT_F_q}) also), ensures
\begin{eqnarray}
\langle \delta F_q(t) \delta F_{q'}(t') \rangle = \frac{k_BT}{Nm_q}e^{-(g_q/m_q) (t-t')} \delta_{q q'} .
\label{force_correlation_q}
\end{eqnarray}
Equation~(\ref{force_correlation_q}) also leads to the correlation of the force fluctuation in accordance with FDT in the GLE (Eq.~(\ref{GLE_v})), i.e., $k_BT \Gamma (t-s)=\left< \delta f (t)\delta f (s) \right>$.
Therefore, a stochastic variable defined as $p(t) = \sum_{q \ge 0} P_q(t) h^{\dagger}_{q,n_0}$, where each mode obeys the dynamics of Eq.~(\ref{Lang_P}) with the noise correlation Eq.~(\ref{FDT_V_q}), exhibits a super-diffusive behavior with $\alpha_p=3/2$.

\subsection{Super-diffusion with tunable exponent}
Following the line of argument for the sub-diffusion (Sec.~\ref{tunable_exponent_subdiffusion}), it is straightforward to control the super-diffusion exponent by introducing the static and dynamical exponents $\nu$ and $z$, which tune the distribution of the mode spectrum. 
Now, Eq.~(\ref{Gamma_power_law_Rouse}) is generalized to
\begin{eqnarray}
\Gamma(t) \simeq k \left|\frac{t}{\tau_u}\right|^{-2/z} 
\label{Gamma_power_law_general}
\end{eqnarray}
which can be decomposed as in Eq.~(\ref{decomp_Gamma}), where the spring and friction constants take the generalized forms given by Eqs.~(\ref{k_q_general}) and~(\ref{gamma_q_general}).
Then one can repeat the subsequent analysis, provided that Eq.~(\ref{k_q_tilde}) is generalized to
\begin{eqnarray}
m_q &=&  k_q^{-1} (q/N)^2 = k^{-1} (q/N)^{1-2\nu} \nonumber \\ 
g_q &=& \gamma_q^{-1} (q/N)^2 = \gamma^{-1} (q/N)^{1+(z-2)\nu}.
\label{k_q_tilde_general}
\end{eqnarray}
From Eq.~(\ref{force_correlation_q}), one can verify the force correlation $\langle \delta f(t) \delta f(0) \rangle \simeq k k_BT (t/\tau_u)^{-2/z}$, which together with the frictional kernel Eq.~(\ref{Gamma_power_law_general}) ensures FDT in the GLE (Eq.~(\ref{GLE_v})).
Therefore, a stochastic variable defined as $p(t) = \sum_{q \ge 0} P_q(t) h^{\dagger}_{q,n_0}$, where each mode obeys the dynamics of Eq.~(\ref{Lang_P}) with the noise correlation Eq.~(\ref{FDT_V_q}), exhibits a super-diffusive behavior with $\alpha_p=2-(2/z)$.

Figure~2 shows plot of the MSD calculated for various exponent $z$ via
\begin{eqnarray}
\langle [\delta(\Delta p_t)]^2 \rangle = k_BT \int_0^t dt_1 \int_0^t  dt_2 \ \Gamma(t_1-t_2),
\label{MSD_from_Gamma}
\end{eqnarray}
where the superposition of mode $q$ was carried out in a discrete form of Eq.~(\ref{decomp_Gamma}) with center-of-mass mode eliminated.
The superdiffusion exponent $\alpha_p = 2-(2/z)>1$ is observed for $z>2$, but the sub-diffusion is not even if $z<2$. 
Being farther away from the applicable range of the integral formula eq.~(\ref{integral_formula}), it looks getting close to the normal diffusion growth \text{$\alpha_p=1$}.

Recalling the argument in \ref{tunable_exponent_subdiffusion}, the mode analyses based on eq.~(\ref{Xq_eq}), \ (\ref{Lang_P}) are mutually complementary. The mode analysis for the super-diffusion does not generate the sub-diffusion, and vice versa.

\begin{figure}[tb]
\begin{center}
\includegraphics[scale=0.50]{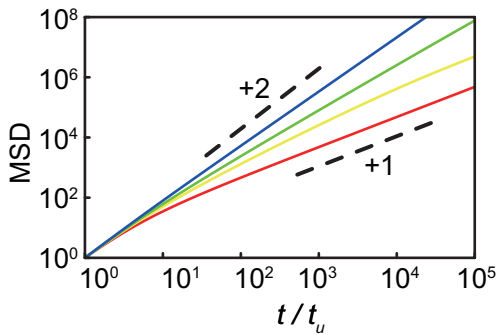}
      \caption{
	  (Color Online) 
	  Plot of MSD calculated according to Eq.~(\ref{MSD_from_Gamma}) for $z=1$ (red), $5/2$ (yellow), $4$ (green), $10$ (blue), where the frictional kernel is given by discrete sum as Eq.~(\ref{decomp_Gamma}). The chain length is $N=10^3$ and the contribution of center-of-mass mode is subtracted.
	  Dashed lines represent the slopes of the upper and lower bounds generated by eq.~(\ref{Lang_P}).
	  All the plots are normalized as $\mathrm{MSD}(t_u)=1$.
	  }
\label{superiff}
\end{center}
\end{figure}

\section{Discussion}
\subsection{Velocity and force correlation}
At the qualitative level, the most essential difference between sub- and super-diffusion processes lies in the correlation functions associated with the kernels.
Equation~(\ref{mu_Rouse}) indicates that, for sub-diffusion process, the velocity at later time is negatively correlated with the earlier velocity, i.e.,  $\langle \delta v(t) \delta v(s) \rangle = k_BT \mu(t-s)<0$ in the time scale $t-s > \tau_u$, and such a negative memory persists for a very long time. In contrast, 
the ``velocity" correlation corresponds to force correlation for super-diffusion process, which is always positive as Eq.~(\ref{decomp_Gamma}) and the FDT $\langle \delta f(t) \delta f(s) \rangle = k_BT \Gamma(t-s)$ show.
Our study pinpoints the corresponding difference in the mode space dynamics, which can be found by comparing Eqs.~(\ref{Xq_eq}) and~(\ref{Lang_P}); while the former has the restoring force $-k_q X_q$, the latter instead has the ``inertial" term $m_q d^2P_q/dt^2$.
Indeed, $X_q$ for $q \geq 1$ spontaneously goes to zero without the external force, but $P_q(t)$ does not even if stopping the operation; instead, the force $F_q(t)=dP_q/dt$ relaxes.

Besides, we add the specific physical interpretations about the contrastive signs between $k_BT \mu (t-s)=\left< \delta v(t) \delta v(s) \right> <0$ and $k_BT \gamma (t-s)=\left< \delta f(t) \delta f(s) \right> >0$
based on the polymer language.
At the constant force, the restoring force inherent to the polymer works against the frictional force.
This indicates that the velocity was faster than the average and the next moment the restoring force acts on the polymer to make the velocity slower than the average. 
Therefore, we see the negative correlation $\left< \delta v(t) \delta v(s) \right> <0$.
On the other hand, at the constant velocity, 
if the force was larger than the average, the deformation gets larger on average in the next moment 
so that the force is needed to get stronger to follow a successive distortion.
This eventually leads to the positive correlation $\left< \delta f(t) \delta f(s) \right> >0$.

\subsection{Polymer dynamics}
So far, we have used the Rouse model (and its generalization) as a tool to analyze the anomalous diffusion processes. It is, however, instructive to discuss the results in terms of polymer physics. 
Here, we provide a physical interpretations on the correspondence between $X_q$ and $P_q$ from viewpoint of the force balance on the cooperatively moving domain, whose size grows due to the tension propagation~\cite{PRE_Sakaue_2007,PRE_Sakaue_2012}.
Let $x(t) = \sqrt{\left< \Delta x_t^2 \right>}$, $f(t)=\sqrt{\left< \Delta f_t^2 \right>}$ denote the characteristic size of the moving domain and the associated force scale. The force balance for the domain reads~\cite{deGennesBook,Macromolecules_Pincus_1976,PRE_Saito_2015}:
\begin{eqnarray}
\gamma_t (t)
\frac{d x(t)}{dt}
\simeq
k_t (t)
x(t),
\label{f_balance}
\end{eqnarray}
where $\gamma_t(t) \simeq \gamma (x(t)/a)^{z-2}$ and $k_t (t) \simeq k_BT/x(t)^2$ is the friction and spring constants, respectively.
Interpreting it as $q$-mode space balance suggests $\gamma_q dX_q/dt = -k_qX_q$.
Adding the external force $\langle F_q(t) \rangle $ and noise $\delta F_q(t)$, where the noise strength imposed on by FDT dictates the spatial fluctuation $X_q$, we arrive at eq.~(\ref{Xq_eq}).
Note that eq.~(\ref{f_balance}) essentially express the fluctuation-dissipation relation $\left< \Delta x_t^2 \right> \simeq \int_0^t ds \ k_BT [\gamma_t (s)]^{-1}$.

The product of the characteristic size of spatial fluctuation and the associated force results in the characteristic energy~\cite{deGennesBook,Macromolecules_Pincus_1976,PRE_Saito_2015};
\begin{eqnarray}
x(t)
\frac{d p(t)}{dt}
\simeq
k_BT
\label{Pincus_2nd}
\end{eqnarray}
where we introduce the characteristic momentum transfer $p(t)=\sqrt{\left< \Delta p_t^2 \right>}$ such that $dp(t)/dt = f(t)$.
Eliminating $x(t)$ with eq.~(\ref{Pincus_2nd}) in eq.~(\ref{f_balance}), we get an alternative expression of the force balance equation
\begin{eqnarray}
\frac{1}{k_t (t)}
\frac{d^2 p(t)}{dt^2}
\simeq
\frac{1}{\gamma_t (t)}
\frac{d p(t)}{dt},
\label{v_balance}
\end{eqnarray}
which is written in the dimension of the velocity.
This suggests the equation of motion in normal mode: $m_q d^2P_q/dt^2 = -g_q dP_q/dt$.
Introducing the enforced velocity and the noise velocity in the same way eventually leads us to eq.~(\ref{Lang_P}).

In most of experiments, the constant velocity protocol would be more accessible than the constant force one.
Measuring $f(t)$ allows us to establish $\left< p(t) \right>$, $\left< [\delta (\Delta p_t)]^2 \right>$ and $\left< \delta f(t) \delta f(0) \right>$. These contain the information corresponding to $\left< x(t) \right>$, $\left< [\delta (\Delta x_t)]^2 \right>$ and $\left< \delta v(t) \delta v(0) \right>$ observed in the constant force protocol.

As an example, let us consider the dynamics of polymer manipulated in the constant force protocol. Close to the equilibrium, the ratio of the fluctuation to the average drift of the tagged monomer is given by the linear response theory
\begin{eqnarray}
\frac{f \left< [\delta (\Delta x_t)]^2 \right>}{k_BT \left< \Delta x_t \right>}=2,
\end{eqnarray}
which is constant with time~\cite{PRE_Sakaue_2013,PRE_Saito_2015}.
If pulled strongly, however, the nonlinear effect generally set in, and we have recently demonstrated the deviation from the above fluctuation-response relation for self-avoiding polymer~\cite{PRE_Saito_2015}. This result was theoretically interpreted using the mode equation for $X_q(t)$, which is a modified version of Eq.~(\ref{Xq_eq}) to include the nonlinear effect in an approximate way.
It would be interesting to perform the corresponding experiment using giant DNA molecule in the constant velocity protocol. For sufficiently small pulling velocity, we should have $v  \left< [\delta (\Delta p_t)]^2 \right>/(k_BT \left< \Delta p_t \right>)=2$, but for large enough velocity, the deviation from it is expected. To investigate such a nonequilibrium fluctuating dynamics in the constant velocity protocol, we expect that the proposed $P_q$ mode analysis Eq.~(\ref{Lang_P}) provides a useful starting point.

Another example is the compression dynamics of DNA in nanochannel~\cite{Reisner2014, Reisner2015}, where the DNA confined in nanochannel is compressed using an optically trapped bead in constant velocity protocol. The dynamics of the force exerted by DNA on the bead can be analyzed, which may provide some useful information on the fluctuation dynamics.

For super-diffusive process, we have $\langle (\delta F_q (t) )^2\rangle =  k_BT /(Nm_q)$ from Eq.~(\ref{force_correlation_q}), as principle of equipartition applied in Eq.~(\ref{Lang_P}) indicates. This leads to $\langle (\delta f(t))^2 \rangle = \sum_q \langle (\delta F_q (t) )^2\rangle \simeq k k_BT$, which is also inferred from Eq.~(\ref{Gamma_power_law_general}) with the approval that $\tau_u$ is the shortest time scale. By introducing the monomer size length scale $a$ such that $k \simeq k_BT/a^2$, the above relation represents the typical order of the force fluctuation $\delta f \simeq k_BT/a$. For sub-diffusion process, a similar reasoning leads to $\langle (\delta v(t))^2 \rangle \simeq k_BT  (\gamma \tau_u)^{-1}$. The characteristic order of the velocity fluctuation is thus evaluated as $\delta v \simeq k_BT/(\gamma a)$.

\section{Summary}
In this article, we have proposed a decomposition of super-diffusive stochastic process $p(t)$ into modes $P_q(t)$, whose relaxation times broadly ranging with a power-law $\sim (N/q)^{\nu z}$. Such a decomposition has been done by referring the well-known decomposition of the sub-diffusive stochastic process $x(t)$ into modes $X_q(t)$ and by exploiting the conjugate relation between $x(t)$ and $p(t)$. 
Our main finding is that the dynamical equation of $P_q(t)$ takes a form of underdamped Langevin equation with the noise velocity $\delta V_q(t)$.
This is contrasted to the well-known form for $X_q$, which is overdamped Langevin equation in harmonic potential with the force noise $\delta F_q(t)$.
The essential difference leads to that the mode analysis for the sub-diffusion does not produce the super-diffusion, and vice versa.

In the context of polymer dynamics, we have pointed out that the way to observe the tagged monomer dynamics has duality. Looking at the force correlation and the super-diffusive behavior of the momentum transfer would be useful to analyze the fluctuating dynamics of a polymer driven by a constant velocity.

Another avenue of the research can be found beyond the MSD analysis. For instance, in ref.~\cite{Amitai}, the first passage statistics of sub-diffusion $x(t)$ process generated
by the superposition of $X_q(t)$ has been investigated. Similar analysis for the super-diffusion $p(t)$ process would be interesting.

Before closing, we have some comments about the scope of the application.
We have focused on the linear polymer as the specific examples, but the theory is not limited to it.
To apply the mode analyses, we have used the assumptions: (i) the independency of modes and (ii) the wavelength-dependent coefficients.
There would be other materials analyzable by this approach as well as the polymerized systems like membranes~\cite{JStatMech_Keesman_Panja_2013,EPL_Mizuochi_2014}.

\section*{Acknowledgement}
We thank Y. Murayama at Tokyo University of Agriculture and Technology and A. Ikeda at University of Tokyo for the fruitful discussion.
This work is supported by KAKENHI (No. 16H00804, ``Fluctuation and Structure") from MEXT, Japan, and JST, PRESTO.

\end{document}